\documentclass[aps,prb,twocolumn,superscriptaddress,showpacs,english]{revtex4-1}

\usepackage[T1]{fontenc}
\usepackage[latin9]{inputenc}
\usepackage{babel}
\usepackage{amsmath}
\usepackage{amssymb}
\usepackage{wasysym}
\usepackage{graphicx}
\usepackage{multirow}
\usepackage[caption=false]{subfig}
\usepackage[dvipsnames]{xcolor}

\usepackage{tikz}
\usetikzlibrary{fadings,shapes.arrows,shadows}   

\tikzfading[name=arrowfading, top color=transparent!0, bottom color=transparent!95]
\tikzset{arrowfill/.style={top color=OrangeRed!20, bottom color=Red, general shadow={fill=black, shadow yshift=-0.8ex, path fading=arrowfading}}}
\tikzset{arrowstyle/.style={draw=BrickRed,arrowfill, single arrow,minimum height=#1, single arrow,
single arrow head extend=.15cm,}}

\newcommand{\tikzfancyarrow}[2][2cm]{\tikz[baseline=-0.5ex]\node [arrowstyle=#1] {#2};}

\newcommand{\Graz}{Institute of Theoretical and Computational Physics, Graz University of Technology, NAWI Graz, 8010 Graz, Austria}
\newcommand{\Vienna}{Faculty of Physics, University of Vienna, 1090 Vienna, Austria}
\newcommand{\Rome}{Department of Physics, Sapienza Universita di Roma, 00185 Rome, Italy}
\newcommand{\RomeInst}{Istituto dei Sistemi Complessi (ISC)-CNR, 00185 Rome, Italy}
\newcommand{\bbo}{BaBiO$_3$}
\newcommand{\sbo}{SrBiO$_3$}
\newcommand{\cbo}{CaBiO$_3$}
\newcommand{\abo}{\textit{A}BiO$_3$}
\newcommand{\monoclinic}{C12\={m}1}
\newcommand{\icubic}{P3\={m}3}
\newcommand{\trigonal}{R3}
\newcommand{\monoclinicS}{P121n1}
\begin{document}

\title{\textit{Ab-initio} study of \abo\ ($A$=Ba, Sr, Ca) under high pressure}

\author{Andriy Smolyanyuk}       \affiliation{\Graz}
\author{Cesare Franchini}        \affiliation{\Vienna}
\author{Lilia Boeri}             \affiliation{\Rome}\affiliation{\RomeInst}

\date{\today}

\begin{abstract}
Using \textit{ab-initio} crystal structure prediction we study the
high-pressure phase diagram of \abo\ bismuthates ($A$=Ba, Sr, Ca) in a pressure range up to 100~GPa.
All compounds show a transition from the low-pressure perovskite
structure to highly distorted, low-symmetry phases at high pressures (PD transition), 
and remain charge disproportionated and insulating up to the highest  pressure studied. 
The PD transition at high pressures in bismuthates can be understood as a combined effect of steric arguments  and of the strong tendency of bismuth to charge-disproportionation. 
In fact, distorted structures permit to achieve a very efficient atomic packing, and at the same time, to have Bi-O bonds of different lengths.
The shift of the PD transition to higher pressures with increasing cation size within the \abo\ series can be explained in terms of chemical pressure.
\end{abstract}


\maketitle

\section{Introduction}
The study of the competition between superconductivity and charge ordering
phenomena has received a strong impulse from recent experiments on high-T$_c$ cuprates\cite{Chang2012} and transition metal dichalcogenides~\cite{kusmartseva_2009}, raising questions
on their interplay and, more in general,
on the role of critical fluctuations in quasi-two-dimensional systems\cite{Snow2003, Dichalcogenides}.

However, the competition between superconductivity and charge density wave (CDW) ordering is not
limited to two-dimensional systems, but has
been observed in other systems with different dimensionality.
A classical example is that of bismuthates with chemical formula \abo, where $A$ is an alkaline earth (Ba, Sr, or Ca).
These compounds are charge-ordered insulators, but undergo an insulator-to-metal transition upon doping, reaching
superconducting $T_c$'s as high as 34~K\cite{Cava1988,Uchida1987,Kazakov1997,Khasanova}. 

The CDW is associated to the mixed-valence behavior of bismuth,
which can acquire two
formal oxidation states --  Bi$^{3+}$ and Bi$^{5+}$ --
which alternate on a perovskite lattice, giving
rise to tilting and breathing distortions.
The charge difference is significantly lower than 2e$^-$ due to the strong hybridization between Bi(s) and O(p) states.
Based on this consideration an alternative picture has been proposed invoking the condensation of holes in lowest O band, resulting in the formation of 
Bi$^{3+}$L$^{2-}$ + Bi$^{3+}$, where L identifies the ligand hole\cite{Sawatzky, Sawatzky2018}.
These two pictures are not mutually exclusive and it is likely that both processes (charge ordering and ligand hole) contributes to the opening of the CDW gap.
Upon doping, the distortion is gradually suppressed and the lattice returns to the ideal perovskite structure\cite{Franchini2009}, which is metallic and superconducting.

Naively, one would expect that extreme pressures could be used to suppress the lattice distortion also in the undoped samples, and hence achieve a metallic, possibly
superconducting state without the complications introduced by doping.
However, as shown in our previous work on \bbo\cite{BaBiO3_Smolyanyuk_PRB_2017}
and independently confirmed by a combined theoretical and experimental work\cite{Martonak2017}, high pressures do not stabilize an ideal perovskite structure,
but a strongly distorted, amorphous-like structure, characterized by strong charge disorder and insulating behavior. 
\begin{figure}[hb!]
 \centering
 \includegraphics[width=\linewidth]{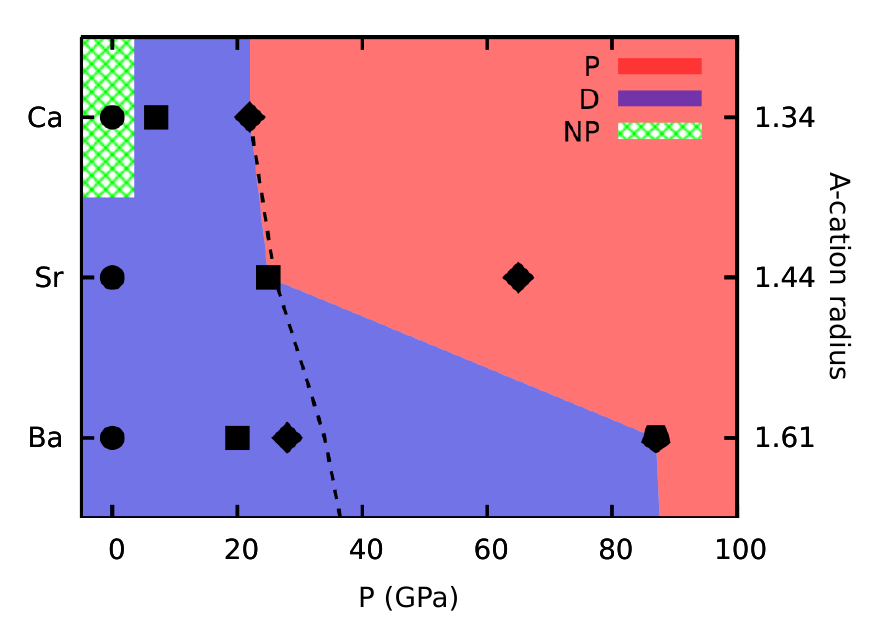}
 \caption{(\textit{color online})
   Phase diagram of the
   structural phase transitions in \abo\ compounds.
   Each transition is labeled by a symbol which represents
   the rank of the structure in the sequence of transitions: \textit{filled circle} for Structure~I, \textit{square}
for Structure~II, \textit{diamond} for Structure~III and \textit{pentagon} for Structure~IV.
``Dark-gray'' (``blue'') region contains perovskite-like~(P)
structures, ``light-gray'' (``red'') -- distorted~(D) structures and
``stroked'' (``green'') non-perovskite (NP) structures.
The thick dashed line indicates the transition line for the perovskite-to-distorted transition (PD transition) estimated from chemical pressure considerations, as explained in Section~\ref{sec:PerovskiteToDistorted}.}
 \label{fig:PhaseDiagram}
\end{figure}

In this work, we investigate this phenomenon further, and address the effect of \textit{chemical} pressure, studying 
the high-pressure phase diagram of the the whole family of \abo\ compounds ($A$=Ca, Sr, Ba) with methods for evolutionary  {\em ab-initio} crystal structure
prediction~\cite{uspex, USPEX_explained}.
The electronic spectra of all predicted phases are computed with
hybrid Hartree-Fock/density functional theory (DFT) functional with the HSE (Heyd-Scuseria-Ernzerhof)
exchange-correlation functional\cite{HSE06}, which has been demonstrated to accurately describe the CDW insulating state of \bbo\cite{Korotin2012,Yin2013,Franchini2009,Franchini2010}, while ordinary DFT describes \bbo\ as a metal.

Our results show that the ordered CDW at low pressures and
the increase charge disproportionation at high pressure, accompanied by larger structural distortions, 
are  general features of the bismuthate phase diagram: in fact, as shown in Fig.~\ref{fig:PhaseDiagram},
all compounds undergo a transition from a perovskite ($P$) to a strongly distorted ($D$) structure, and remain insulating up to 100 GPa.
The occurrence of distorted structures is a consequence of a steric argument,
combined with the tendency of bismuth to charge disproportionation.

This paper is organized as follows: in Section \ref{sec:Results} we describe the results of our \textit{ab-initio} calculations starting with our prediction for
the high-pressure phase diagram  in Subsection~\ref{subsec:PredictedPhaseDiagram}, followed by the description of electronic properties of the most stable structures in Subsection~\ref{subsec:ElectronicStructure}.
The discussion on the perovskite-to-distorted transition (PD transition) follows in Section~\ref{sec:PerovskiteToDistorted}.
The main conclusions of this work are summarized in Section~\ref{sec:Conclusions}.
Computational details are described at the end of the paper in Section~\ref{sec:ComputationalDetails}.

\section{Results}
\label{sec:Results}


\subsection{Phase Diagram}
\label{subsec:PredictedPhaseDiagram}

In Fig.~\ref{fig:PhaseDiagram} we show the  phase diagram of \abo\ in the pressure range from 0 to 100~GPa, predicted using {\em ab-initio} evolutionary
crystal structure prediction methods.
For Bismuthates, the use of an unbiased crystal structure prediction method
is crucial,  since the group-subgroup approach\cite{Howard,interplay_2013,Structural_determination, He} which is routinely used to predict structural transitions in perovskites is not able to capture the transition to the distorted structure
-- see Ref.~\onlinecite{BaBiO3_Smolyanyuk_PRB_2017} for details.

To generate the phase diagram in Fig.~\ref{fig:PhaseDiagram},
a set of reference crystal structures were first obtained from evolutionary crystal structure prediction runs at 0, 50 and 100~GPa.
The most promising structures from the pool of crystal structures obtained at these pressures are further relaxed at intermediate pressures at constant volume intervals, and interpolated with an analytical equation of state, which allowed us to determine the transition pressures accurately
(see Sec.~\ref{sec:ComputationalDetails} for more details).
The lowest-enthalpy structures at each pressure were then used to construct
the final phase diagram.

We have plotted all \abo\ compounds on the same figure (Fig.~\ref{fig:PhaseDiagram}), with a common pressure axis;
the compounds are equispaced along the 
vertical axis; starting with Ca, which has the smallest ionic radius, 
the size of the $A$ cation increases along the $y$ axis from Ca to Ba
exerting an effective {\em chemical pressure} on the perovskite lattice 
(see Table~\ref{table:volumes} for corresponding ionic radii and unit cell
volumes).
Assuming that the most important parameter governing the structural
transitions is the volume of the unit cell, and that this is mainly
determined by the size of the $A$ cation, this means that the sequence of structural transitions seen in \bbo\ should occur at lower pressures in \sbo\ and even lower in \cbo.

In the figure, each transition is labeled by a symbol representing
 the rank of the structure in the transition series.
 The diagram is also divided into three regions, denoted with the letters
 P (perovskites), D (distorted) and NP (non-perovskite).
The thick dashed line indicates the shift of perovskite-to-distorted transition pressure (PD transition), estimated from the simple cubic perovskite model, as explained in Section~\ref{sec:PerovskiteToDistorted}.

We start from the low-pressure region.
An empirical guess of the ambient pressure structure for each compound
can be obtained from the Goldschmidt tolerance factor, which is based on the size mismatch between the cations. 
The tolerance factor for double perovskites is defined as:
$t=\frac{r_A+r_O}{\sqrt{2} (<r_{Bi}> + r_O)}$, where
$r_A$ is the radius of the $A$ cation, $r_{Bi}$ is averaged ionic radius of bismuth and $r_O$ is radius of oxygen. 
The ideal value $t=1$ corresponds to a cubic perovskite structure;
deviations from this ideal value indicate the amount of distortion
needed to stabilize the atomic arrangement:
 the higher the deviation, the higher the distortion.
Based on the value of $t$, one would expect that  at ambient pressure all \abo\ compounds should be either monoclinic or orthorhombic perovskites
with tilted oxygen octahedra, because $t<0.97$\cite{TolFactor}
for all $A$ cations (see Table~\ref{table:volumes}).
Indeed, the tolerance factor correctly predicts the structure at  ambient pressure for \bbo\ and \sbo, which are known experimentally\cite{COX1976, Kazakov1997}.
However, the structure obtained by our evolutionary algorithm
predictions for \cbo\ is trigonal, which apparently contradicts
the argument based on the tolerance factor. Note,
however, that in this case the value of $t$ is extremely small,
and several exceptions to the tolerance factor argument are known\cite{PerovkitesReview}.

\begin{table}
 \begin{center}
  \caption{\label{table:volumes}
Ion Shanon radii (R) for the \textit{A} cation in coordination XII;
corresponding tolerance factors ($t$) for
\abo\ compounds in the double perovskite structure;
calculated volume per formula unit ($V$),
average Bi-O distance ($\overline{BiO}$ ) and
breathing distortion $\delta$ for \abo\ at P=0~GPa.
$\delta=\frac{1}{2}(\overline{Bi_1O}-\overline{Bi_2O}$), where $\overline{Bi_1O}$ and $\overline{Bi_2O}$
are average Bi$_1$-O and Bi$_2$-O bond distances respectively.
}
\begin{ruledtabular}
  \begin{tabular}{cccccc} 
     & R(A) (\r{A}) & V (\r{A}$^3$/f.u.) & t    & $\overline{BiO}$ (\r{A}) &  $\delta$ (\r{A}) \\ \hline
\cbo & 1.34         & 76.91              & 0.85 & 2.29 & 0.122 \\
\sbo & 1.44         & 80.00              & 0.88 & 2.25 & 0.089 \\
\bbo & 1.61         & 85.87              & 0.93 & 2.24 & 0.074 \\
\end{tabular}
\end{ruledtabular}

 \end{center}
\end{table}

\begin{figure}[b]
 \centering
 {\scriptsize\tikzfancyarrow[0.9\linewidth]{\textit{A}-cation radius}}
 
   \begin{subfloat}[\cbo\ I(NP) \label{fig:CBO}]{   
   \includegraphics[height=64\unitlength]{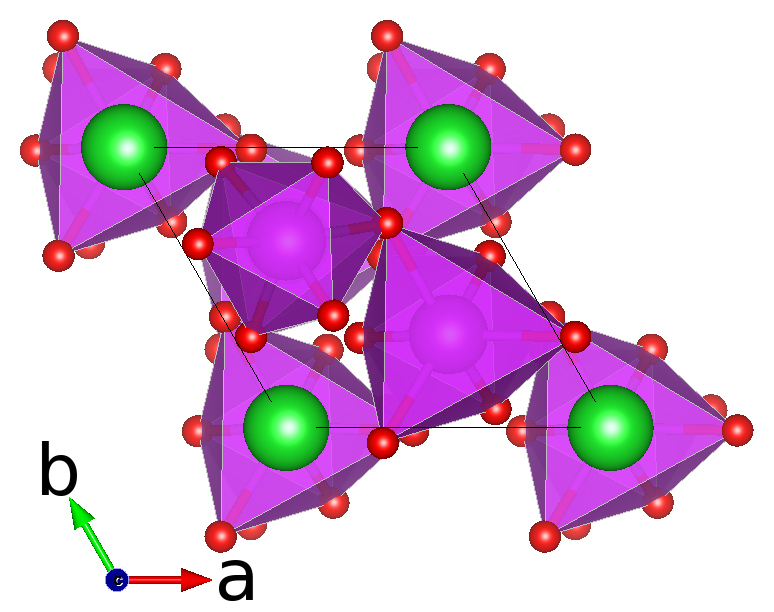}}
   \end{subfloat}%
   \begin{subfloat}[\sbo\ I(P) \label{fig:SBO}]{
   \includegraphics[height=64\unitlength]{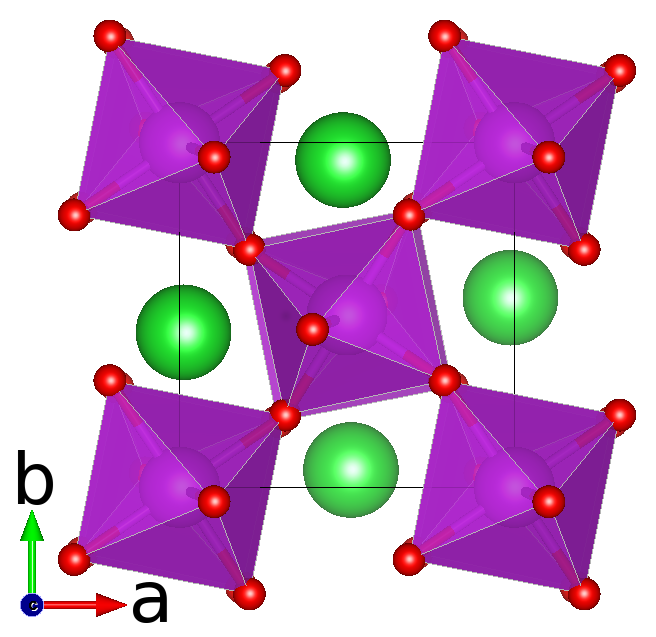}}
   \end{subfloat}%
   \begin{subfloat}[\bbo\ I(P) \label{fig:BBO}]{
   \includegraphics[height=64\unitlength]{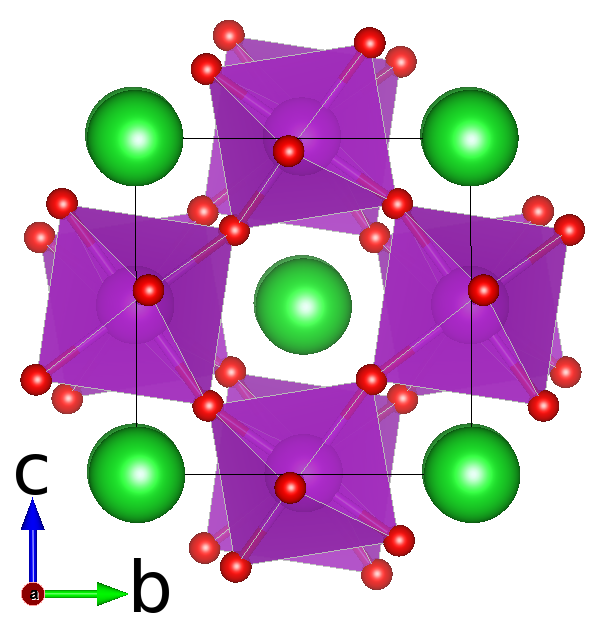}}
   \end{subfloat}

   \begin{subfloat}[\cbo\ III(D) \label{fig:CBOD}]{   
   \includegraphics[height=60\unitlength]{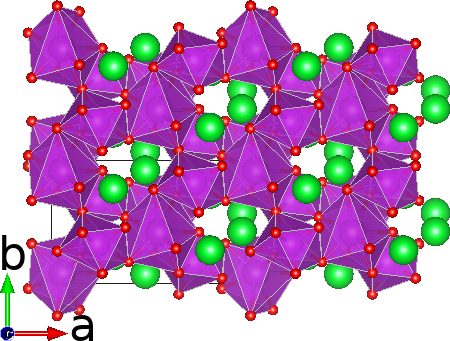}}
   \end{subfloat}%
   \begin{subfloat}[\sbo\ III(D) \label{fig:SBOD}]{
   \includegraphics[height=60\unitlength]{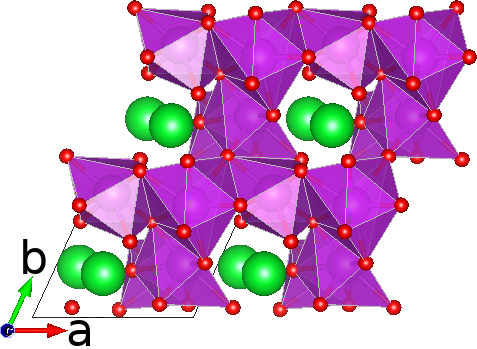}}
   \end{subfloat}%
   \begin{subfloat}[\bbo\ IV(D) \label{fig:BBOD}]{
   \includegraphics[height=60\unitlength]{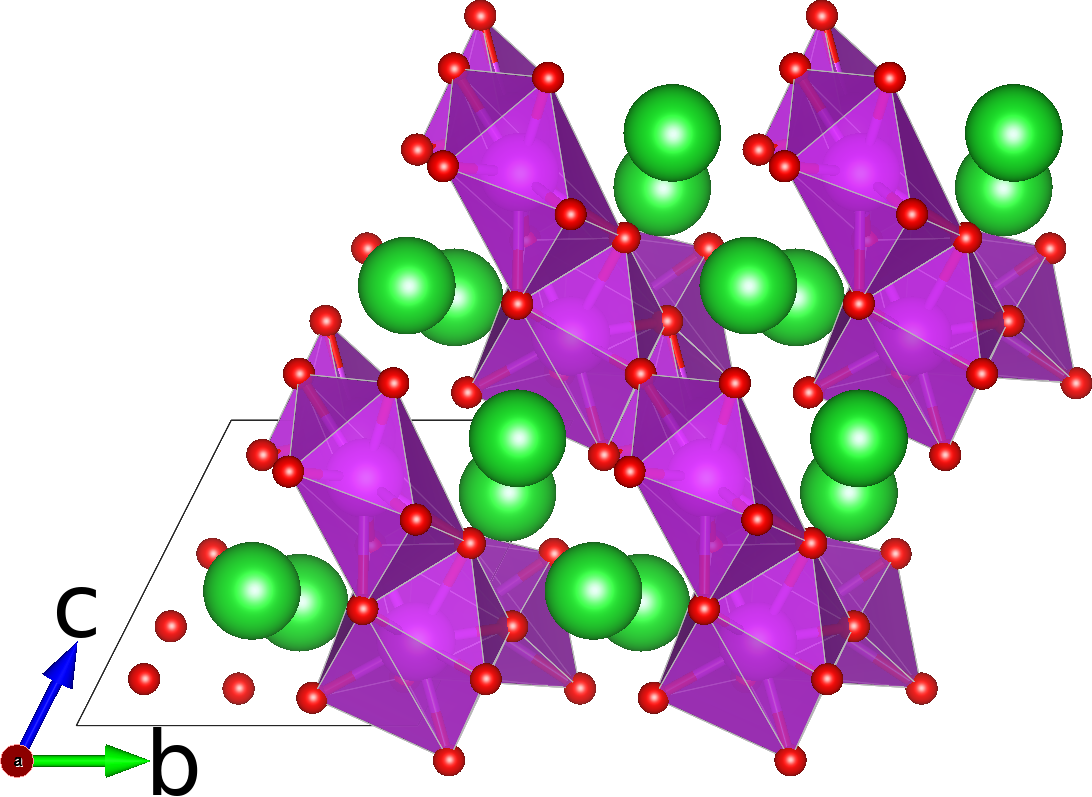}}
   \end{subfloat}  
   
    \caption{({\em color online})
 Structures of \abo\ compounds used in this work:
    (a)~non-perovskite \cbo (), perovskite
    (b)~\sbo,
    (c)~\bbo\ and  distorted
    (d)~\cbo,
    (e)~\sbo\ and (f)~\bbo.
 $A$ cations (Ba, Sr, Ca) are shown as light-gray (light-green) spheres, Bi -- (gray) purple  spheres inside  dark-gray (purple) polyhedron, O -- small gray (red) spheres.}%
    \label{fig:ABO_structures}
\end{figure}

At ambient pressure \bbo\ and \sbo\ structures exhibit different monoclinic
symmetries: C12\={m}1 and P2$_1$/c respectively
(see Fig.~\ref{fig:BBO} and \ref{fig:SBO}). Both structures contain the characteristic perovskite building block formed by BiO$_6$ octahedra and  share a common parent structure -- an ideal cubic perovskite with \icubic\ symmetry.
The two monoclinic structures can be obtained applying two different types
of distortions to the parent structure: \textit{breathing} (alternating the size of Bi-O octahedral environment) and \textit{tilting} (rigid rotations of Bi-O octahedra). The difference between the two compounds is the different tilting pattern ($a^0b^-b^-$ and $a^+b^-b^-$ for \bbo\ and \sbo\ in Glazer notation\cite{Glazer}, respectively) and the amount of breathing distortion (0.074~\AA\ and 0.089~\AA, see Table~\ref{table:volumes}).

Upon increasing pressure we observe that all compounds undergo a series of
structural transitions.
The common feature  is that all \abo\ compounds undergo a transition to a highly distorted phase~(D) beyond some critical pressure.
By highly distorted phase we mean a structure
that does not belong to the perovskite family,
in the sense that  a transition from the perovskite structure
requires large atomic displacements and strong distortions of
the original structural motifs, which cannot be  decomposed into breathing and tilting distortions\cite{comment}. Even though they do not exhibit the regular arrangement seen at ambient
 pressures, the Bi sites in the high-pressure distorted structures
 exhibit charge disproportionation and insulating behavior, as discussed
 in Sect.~\ref{subsec:ElectronicStructure}.

 Distorted structures in the bismuthates were predicted for the
 first time by us in \bbo\cite{BaBiO3_Smolyanyuk_PRB_2017}, and later
 confirmed by Ref.~\onlinecite{Martonak2017}.

We now describe the phase diagrams of the individual compounds in detail;
the corresponding structural files are provided in Supplementary Materials.

\textit{\bbo:} The data for \bbo\ presented in this work are taken from our previous work~\cite{BaBiO3_Smolyanyuk_PRB_2017}.
We find that a first transition from the monoclinic I(P) (space group \monoclinic; see Fig.~\ref{fig:BBO}) 
to triclinic II(P) structure occurs at 20~GPa.
The II(P) structure is another form of perovskite structure with an additional octahedral tilting axis compared to the I(P)
structure.
The next transition is from the triclinic II(P) to a \textit{clustered} monoclinic III(P) structure at about 28~GPa.
This structure can be described as a perovskite with the stacking fault resulting in only half of the Bi atoms exhibiting an octahedral environment.
Although it strongly deviates from the perovskite structure, we still assign it to the perovskite family (P). 
The last transition is from a \textit{clustered} monoclinic to a distorted IV(D) (see Fig.~\ref{fig:BBOD})
structure at 87~GPa, which has no symmetry.

\textit{\sbo:} The first transition is from a perovskite-like monoclinic I(P) (space group \monoclinicS; see Fig.~\ref{fig:SBO})
to a \textit{distorted} II(D) structure at about 25~GPa (see Fig.~\ref{fig:SBOD}) and the second is from a distorted II(D)
to another distorted structure III(D) at 65~GPa.
These are two different \textit{distorted} structures,
but is difficult to  give a clear description of the structural differences
as the structures are too distorted to identify a clear pattern.

\textit{\cbo:}
There are no experimental crystal structure data for \cbo\ at  ambient pressure,
but only a computational study suggesting a polar R3 structure\cite{He}.
Indeed, \cbo\ is unstable at ambient conditions but it is possible to stabilize it
at high pressures\cite{CBOsynthesis} (about 6~GPa), or
upon $K$ doping,
Ca$_{x}$K$_{1-x}$O$_3$, with $0.15\leq x\leq 0.25$\cite{Khasanova}.
According to our evolutionary predictions the structure at ambient pressure is trigonal (\trigonal)
and consists of ``paired'' distorted octahedra that share a common edge.
We find that the Bi-Bi and Ca-Ca environment is no longer octahedral, as in perovskites, but tetrahedral.
Therefore, in Fig.~\ref{fig:PhaseDiagram} this structure
is classified as a non-perovskite structure ($NP$).
Our structure is  12~meV/atom (17~meV/atom in HSE) lower
in enthalpy than the perovskite structure proposed by {\em He et al.} in
Ref.~\onlinecite{He} and based on the traditional
group-subgroup symmetry analysis\cite{Howard}.

The first structural transition is from the trigonal I(NP) (see Fig.~\ref{fig:CBO})
to monoclinic II(P) structure at 7~GPa.
In this case the Bi-Bi and Ca-Ca environment is distorted octahedral and the Bi-O octahedra are distorted too, with additional tilting.
We therefore classify this structure as perovskite (P).
The second transition is from the II(P) to a distorted III(D) (see Fig.~\ref{fig:CBOD})
structure with C1c1 symmetry at 22~GPa.

In summary, our study of the phase diagram of \abo\ compounds shows
that pressure does not stabilize more symmetric structures,
as in other transition metals perovskites, but highly distorted ones,
as we originally observed for  \bbo\cite{BaBiO3_Smolyanyuk_PRB_2017}.
 Furthermore, reducing the size of the $A$ cation shifts the transition
from perovskite-like 
 to distorted structures to a lower pressure, in agreement with
 considerations based on {\em chemical pressure},
 i.e. with the idea that substituting a cation with a smaller one is
 equivalent to applying an external pressure.
In the next sections, we will study in more detail the nature and origin
of the PD transition and the electronic properties of the most relevant structures.

\subsection{Electronic Structure}
\label{subsec:ElectronicStructure}
     
Fig.~\ref{fig:DOS} shows the  total and atom-projected density of states (DOS) for all \abo\ compounds,
at the transition pressure between the perovskite~(P) and the distorted~(D) structures,
obtained with the HSE functional.
For consistency, also for \bbo\ we show the DOS for the perovskite and the
distorted structures, at the transition pressure between the two, but
we remind that in this region the stable structure is the {\em clustered}
one. The aim of the figure is in fact to trace the origin
of the PD transition, which is common to all \abo\ compounds,
while the {\em clustered} region is found only in \bbo.

As it can be clearly seen, all structures are insulating, with a gap at ambient pressure:
0.6~eV for \bbo\ (experimental measurements vary from 0.2~eV\cite{SLEIGHT197527}
to 1.1~eV\cite{KUNC1991325}),
1.1~eV for \sbo\ and 2.6~eV for \cbo.
The increase of the gap with decreasing cation size is consistent with
the larger breathing distortions predicted for the \cbo\ structure
(see Table~\ref{table:volumes}).
The behaviour of the gap as a function of pressure
is shown in Fig.~\ref{fig:Gaps}\cite{clustered}.

The mechanism leading to the opening of a gap in the perovskite structure
has been discussed by several authors: charge disproportionation at the
Bi site causes alternating breathing distortion of the
ideal perovskite lattice, further stabilized by tilting distortion, and this opens a gap in the strongly hybridized Bi(s)-O(p) antibonding band\cite{Franchini2010, Sawatzky}.
This is evidenced by the partial DOS plots, which show a
clear splitting and charge redistribution between Bi$^{3+}$ and Bi$^{5+}$
states, localized below and above the semiconducting gap.
A similar situation is found in distorted phases, where it is still possible
to identify inequivalent Bi$^{3+}$-- and Bi$^{5+}$--like ions.
One clearly sees that the two contributions to the DOS
are well separated, with Bi$^{5+}$ and  Bi$^{3+}$ states
giving the dominant contribution to the electronic DOS
above and below the gaps, respectively.
In distorted \sbo\ and \bbo\ the two main peaks are further split into several
sub-peaks, indicating a more complex pattern of charge disproportionation.

In typical transition metal perovskites the application of pressure leads to a
continuous closing of the gap\cite{PhysRevB_85_195135}.
However, for Bismuthates the situation is different: increasing pressure by chemical substitution makes the gap larger, due to the gradual narrowing of the Bi$^{3+}$ and Bi$^{5+}$ peaks associated with the upward shift and broadening of the occupied Oxygen band.
Despite the larger degree of Bi-O hybridization the narrowing of the Bi$^{3+}$ and Bi$^{5+}$ peaks increases the bonding-antibonding repulsion,
increases the bond- (see Table~\ref{table:volumes}) and
charge-disproportiation (see Fig.~\ref{fig:Charges}) and ultimately lead to a substantial increase of the gap size.

As already mentioned, the classification into  Bi$^{3+}$ and Bi$^{5+}$ is clearly only indicative:
due to the strong hybridization with oxygen the actual charge disproportionation
between Bi sites is much smaller than two. In order to obtain a
quantitative estimate, we performed a Bader charge analysis\cite{Bader,web_bader, baderpaper,bader_comment}, which is more
accurate than the estimation based on the integration of the DOS, that we
used in our previous works\cite{BaBiO3_Smolyanyuk_PRB_2017, Franchini2009}.
Remarkably, we found that Bader charge analysis deliver charge differences
between Bi$^{3+}$ and Bi$^{5+}$ ions 3-4 times larger.
The results are shown in Fig.~\ref{fig:Charges}, where
$\Delta\rho$ is a difference between the maximum and minimum charge located on bismuth atoms obtained in the Bader analysis.
Each symbol in the figure represents a specific structure and its filling shows if the structure is perovskite-like
-- filled symbol --  or distorted --  empty symbol.
The figure shows that the average charge difference is of $\sim 0.5~e^{-}$/f.u., and slightly increases with pressure; apart from \bbo, where the
presence of the clustered structure complicates the phase diagram,
the transition from the perovskite to the distorted structure is relatively
smooth, similarly to what we observed for the DOS.

Having clarified the reasons behind the robustness of the CDW gap upon
pressure we move now to the analysis of the PD structural transition,
which is done in the following section.

\section{Perovskite-to-Distorted Transition}
\label{sec:PerovskiteToDistorted}

The first obvious argument to explain the PD transition is a
\textit{steric} one, i.e. distorted structures, which are
very compact, become favorable at high pressures,
where the $pV$ term becomes the dominant part of the total enthalpy.
In fact, at the transition pressures between perovskite
and distorted structures,
the volumes of the distorted structures are on average $\sim 7.5 \%$
smaller than those of the perovskite ones.

The lower specific volume of the distorted structures is achieved
by a more efficient packing of the atoms with an increasing coordination number.
To visualize this, in  Fig.~\ref{fig:Neighbors} we plot the average number of oxygens in bismuth environment  as a function of pressure.
Values close to six represent perovskite or perovskite-like structures.
The higher the difference from this value, the higher the amount of distortion in the structure.

Clearly, all \abo\ compounds undergo a discontinuous jump from perovskite-like structures to distorted ones at different critical pressures: \cbo\ and \sbo\ at 22-25~GPa; and \bbo\ at 87~GPa (45~GPa if the \textit{clustered} structure is neglected).
  Thus, the increase of coordination number is the general mechanism that
  stabilizes distorted structures at extremely high pressures in agreement with what we found for \bbo\ in the previous paper\cite{BaBiO3_Smolyanyuk_PRB_2017}.

  The origin of this effect can be understood as follows:
    initially, pressure causes a compression of the Bi-O bonds.
  However, the compressibility of Bi-O bonds  
  decreases with increasing pressure, until it reaches a
  critical value, where it becomes energetically more
  favorable to introduce new chemical bonds 
and rearrange bond distances than to continue to compress bonds.

This can be easily seen from Figure~\ref{fig:BiO_length},
showing the pressure-dependence of the average Bi-O length;
for all \abo\ compounds there is a clear jump to larger Bi-O
length  at the pressures corresponding to the transition from perovskite-like to distorted structure.
The increase in coordination number, and the tendency
to form disordered structures, is also favored by
the intrinsic tendency of bismuth to charge disproportionation,
which leads to forming two (or even more) types of bonds with different lengths.
The tendency to charge disproportionation
increases with increasing pressure, as shown
by the Bader charge analysis in Fig.~\ref{fig:Charges},
and hence the arrangement in high-symmetry structures
becomes progressively less favorable.

Indeed, forming {\em distorted} structures, which have a very
low symmetry,
is an efficient way to increase the number of bonds
in a very closely-packed structure.
Let us consider first the Bi-O octahedral environment:
the Bi-O coordination geometry is restricted by the chemical composition and
crystallographic restriction theorem, which forbids
Bi-O to form a regular polyhedron 
with coordination number larger than six.
For example, the 7-coordinated pentagonal dipyramid has a 5-fold rotational symmetry and the 8-coordinated square antiprism has an 8-fold symmetry, which are both forbidden due to the crystal periodicity. Cubic coordination is not allowed by chemical composition as it requires a composition with the same number of Bi and O.
As result, when the coordination number increases due to the formation
of new bonds under increasing pressure, pronounced distortions of the local environment occur, that lead to highly-distorted structures at high pressure.

The perovskite-to-distorted transition pressure increases with
increasing cation radius (see Fig.~\ref{fig:PhaseDiagram}).
This effect, sometimes referred to as {\em chemical pressure},
is based on the idea that replacing a cation
with a larger one, keeping the volume fixed,
increases the effective pressure on the BiO$_3$ sublattice.
This effect is easily illustrated by a simple numerical experiment, based
on the ideal cubic perovskite structure.

\begin{figure}
 \centering
 \includegraphics[width=\linewidth]{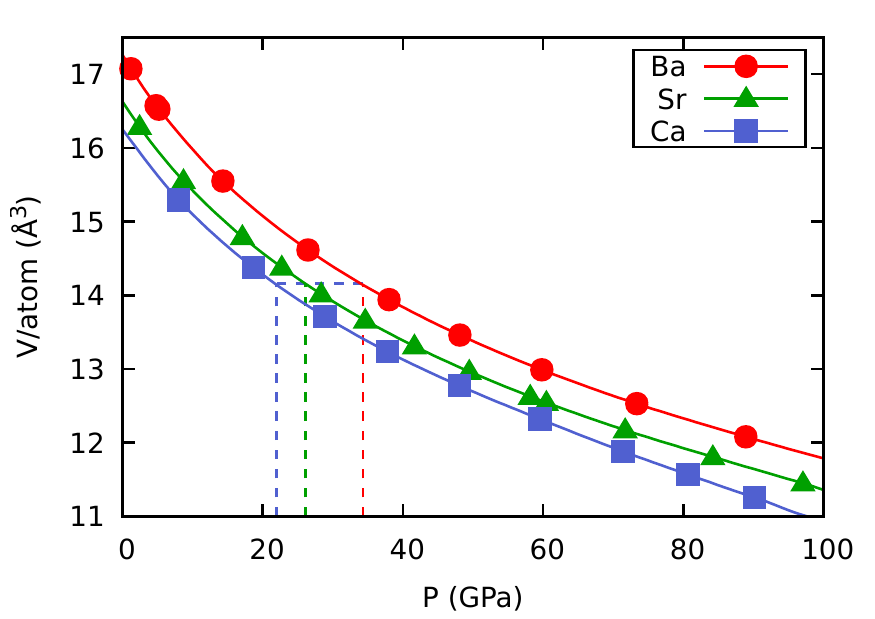}
 \caption{(\textit{color online})
    Equation of state for \abo\ compounds in the ideal cubic perovskite
   structure, used as a model to estimate the shift of the PD transition line due to the chemical pressure,
   showed by dashed lines in Fig.~\ref{fig:PhaseDiagram}. The PD transition at 22~GPa for \cbo\ is taken as a reference point
   to estimate a common volume for the transition (see text).
 Lines are meant as a guidance for the eye.}
 \label{fig:Chemical_pressure}
\end{figure}

Fig.~\ref{fig:Chemical_pressure} shows the equation of state
(EOS) for  \abo\ compounds in cubic structure.
The influence of \textit{chemical} pressure may be estimated
calculating the pressures, corresponding to the intersection of
a $V=const$ line with the EOS curve for a given compound.
For example, according to Fig.~\ref{fig:PhaseDiagram}, the PD transition pressure for \cbo\ is at around 22~GPa.
We can use this value and the EOS for our cubic model to estimate the PD transition pressures in \sbo\ and \bbo. First, we find the volume, which corresponds to 22~GPa on the EOS for cubic \cbo\ ($V_{PD}$).
If we assume that the main driving force of the PD transition is the
unit cell volume, we can estimate the PD transition pressure for \sbo\
and \bbo\ from the intersection of the $V=V_{PD}$ line with the
relative EOS curves. The two values are shown dashed lines
on the Fig.~\ref{fig:Chemical_pressure}.
Our crude estimate gives 26~GPa for \sbo\ and 34~GPa for \bbo;
the agreement with the transition pressure is remarkable
for \sbo\ (25~GPa), but apparently quite bad for \bbo\ (87~GPa).
However, we recall that in \bbo\ a \textit{clustered} phase,
in which only half of the Bi atoms are arranged in a regular octahedral environment, is predicted to occur in-between the perovskite and the distorted structure, shifting the transition to fully distorted structures to 87 GPa.
If one disregards the \textit{clustered} phase, the transition from perovskite to distorted structure occurs at around 45~GPa, which is
in fair agreement to the 34~GPa value, predicted from the ideal cubic model.

\section{Conclusions}
\label{sec:Conclusions}
In conclusion, in this work we studied the pressure-behaviour
of three rare-earth bismutathes \abo\ ($A$=Ba, Sr, Ca),
using {\em ab-initio} crystal structure prediction
and hybrid functional calculations.
Similarly to what previously observed in \bbo, we find
that charge disproportionation is robust and not suppressed by pressure in all \abo\ compounds, which also all remain insulators up to 100~GPa.
The charge disproportionation between Bi$^{3+}$ and Bi$^{5+}$, estimated
by Bader charge analysis, is much higher than what predicted
by usual methods based on the DOS, starting from around 0.5 $e^{-}$ at ambient pressure and going up to 0.7 $e^{-}$ for the highly distorted high-pressure
structures.

Indeed, the tendency to charge disproportionation becomes stronger with increasing pressure and has a big influence on the crystal structure; in fact,
all compounds undergo a transition from perovskite structure to a highly distorted structure (PD transition); in \bbo, this happens through a transition to an
intermediate \textit{clustered} phase, whose presence shifts the
PD transition to higher pressures, with respect to the value estimated
on the basis of chemical pressure.

The formation of distorted structures is explained by a steric effect,
combined with the mixed-valence behavior of bismuth;
distorted structures have a higher density (more efficient packing) than
any variant of the perovskite structure, and this permits
to reduce the $pV$ contribution to the enthalpy which is dominant at high pressures. In the distorted structures, a more efficient packing is achieved by
allowing the coordination number to increase beyond the value $6$ for the
characteristic of the perovskite structure; this tendency,
combined with that to  charge disproportionation,
means that it is not possible to form any symmetric structure under this circumstances.

\section{Computational Details}
\label{sec:ComputationalDetails}
To construct the high-pressure phase diagram,
candidate structures with 10, 20 and 40 atoms per unit cell at 0~GPa, 50~GPa and 100~GPa were generated using the \texttt{USPEX} code
for evolutionary crystal structure prediction.~\cite{uspex,uspex2,uspex3}.
For each group of structures, generated by a separate \texttt{USPEX} run,
we chose 5 lowest in enthalpy to perform a final accurate structure relaxation.
The structure relaxations were performed allowing atoms relaxations and change of the cell shape,
but the volume was fixed.
Each structure was than relaxed on a grid of at least 9 volumes,
chosen between the average volumes corresponding to 0 and 100 GPa.
The pressures of the relaxed structures were estimated from the stress tensor.
The resulting pressure vs volume relation were then interpolated analytically;
from the interpolation curves, we obtained the transition pressures used to
derive the final phase diagram shown in Fig.~\ref{fig:PhaseDiagram}.

\begin{figure}
 \centering
 \includegraphics[width=\linewidth]{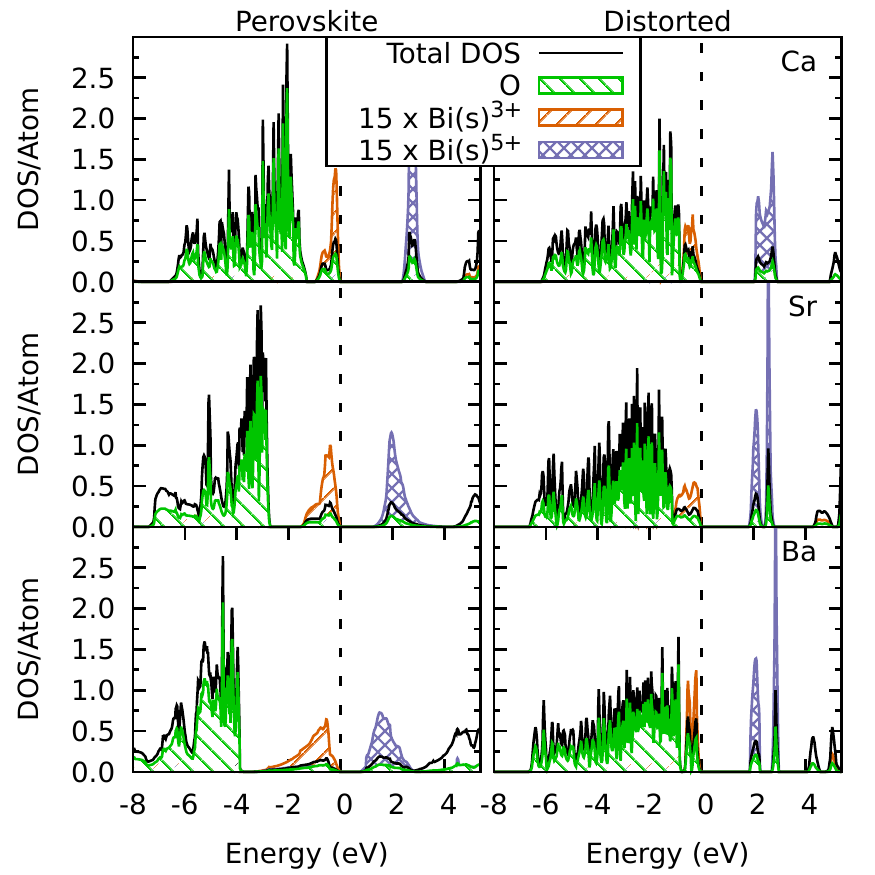}
 \caption{(\textit{color online}) Total and atom-projected DOS
   for \abo\ compounds before and after the PD transition: 
   22~GPa for \cbo, 25~GPa for \sbo\ and 45~GPa for \bbo,
   calculated with the HSE functional; the
    structures were optimized at the PBE level.
Bi atoms are divided into formal Bi$^{3+}$ and Bi$^5+$ valences.
For the sake of clarity the Bi-DOS is multiplied by a factor 15.}
 \label{fig:DOS}
\end{figure}

\begin{figure}
 \centering
 \includegraphics[width=\linewidth]{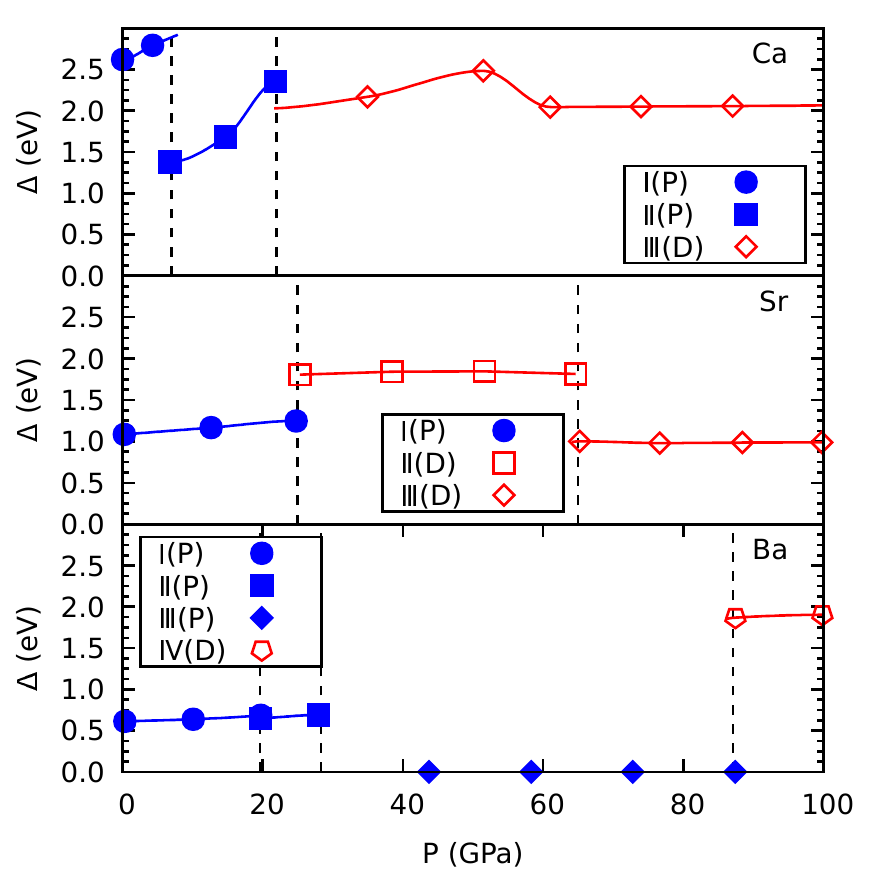}
 \caption{(\textit{color online}) Band gaps for the most stable
   structures of \abo\ compounds as a function of pressure\cite{clustered}.
   Each symbol represents a specific structure, as defined in
   the caption of Fig.~\ref{fig:PhaseDiagram};
   the filling indicates whether
   the structure is perovskite-like (filled symbol)
 or distorted (empty symbol). Lines are meant as a guidance to the eye.}
 \label{fig:Gaps}
\end{figure}

\begin{figure}
 \centering
 \includegraphics[width=\linewidth]{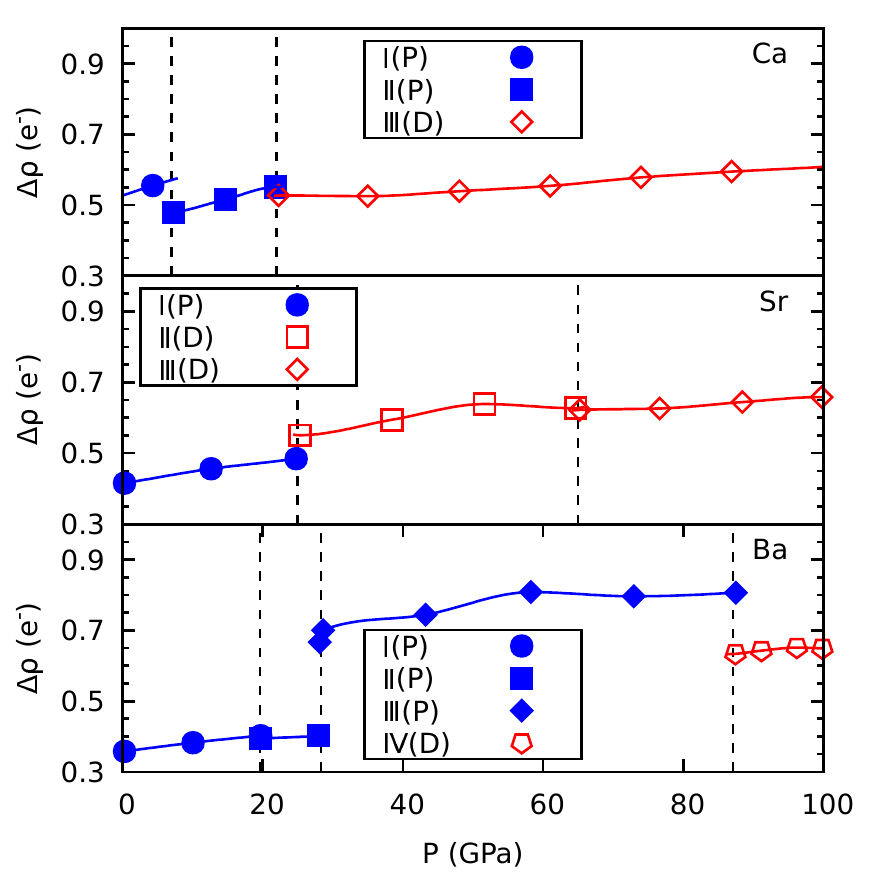}
 \caption{(\textit{color online}) Difference between maximum and minimum charge ($\Delta \rho$) located on bismuth atoms w.r.t pressure,
   obtained using Bader analysis.
   Note that the Bader charge analysis yields sensibly higher values than
   the estimate based on  the partial DOS estimate used in our previous paper\cite{BaBiO3_Smolyanyuk_PRB_2017}.
   The meaning of the symbols is the same as in Fig.~\ref{fig:Gaps}.}
 \label{fig:Charges}
\end{figure}

\begin{figure}
 \centering
 \includegraphics[width=\linewidth]{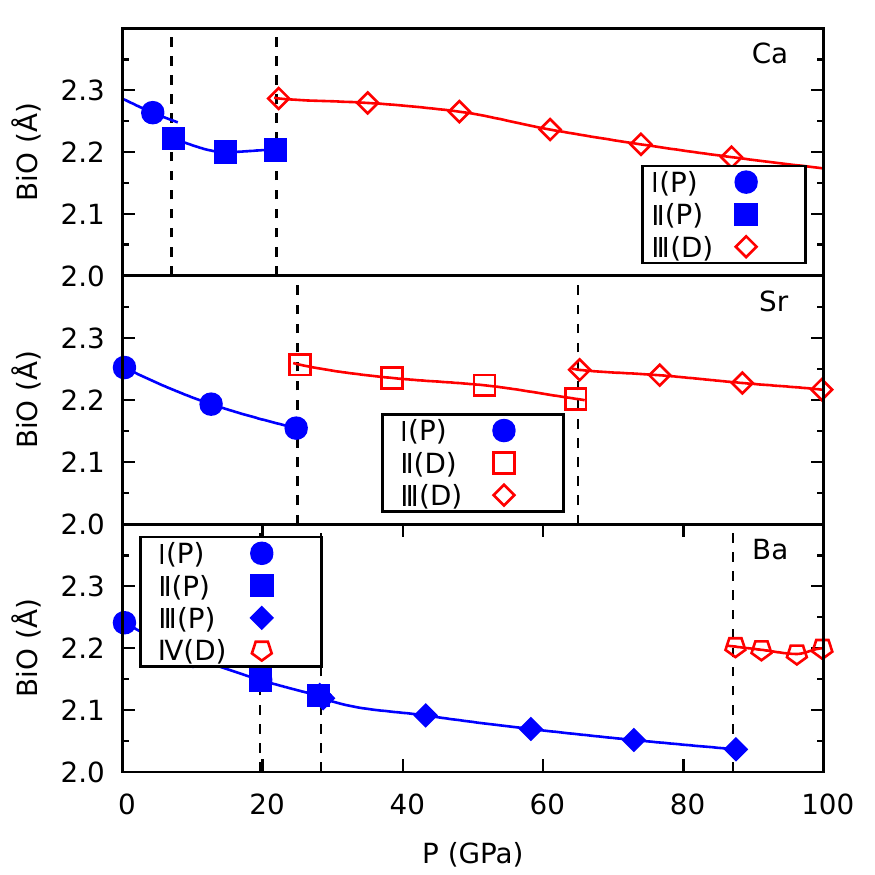}
 \caption{(\textit{color online}) Average BiO bond length as a function of pressure.
    The meaning of the symbols is the same as in Fig.~\ref{fig:Gaps}.}
 \label{fig:BiO_length}
\end{figure}

We used density functional theory (DFT) in the
generalized gradient approximation (GGA)
with Perdew-Burke-Ernzerhof functional\cite{GGA1, GGA2}
to calculate the total energies and perform structural optimization, 
as implemented in the \texttt{VASP} package \cite{VASP1, VASP2, VASP3, VASP4} 
using projector augmented wave method (PAW) pseudopotentials \cite{PAW1, PAW2}.
The hybrid HSE functional (Heyd-Scuseria-Ernzerhof) was used
only to compute electronic spectra, in particular for the DOS calculations (Fig.~\ref{fig:DOS}) and for calculation of band gaps (Fig.~\ref{fig:Gaps})\cite{clustered}.
The energy cutoff value was set to 500~eV and $\Gamma$-centered Monkhorst-Pack grid \cite{MP1, MP2} with
the reciprocal-space resolution 0.04~$2\pi \mbox{\r{A}}^{-1}$
was used for the GGA for structural relaxation and 0.01~$2\pi \mbox{\r{A}}^{-1}$ for total energy calculation.
For calculations where HSE functional was used the reciprocal-space resolution was 0.04~$2\pi \mbox{\r{A}}^{-1}$.

\begin{figure}
 \centering
 \includegraphics[width=\linewidth]{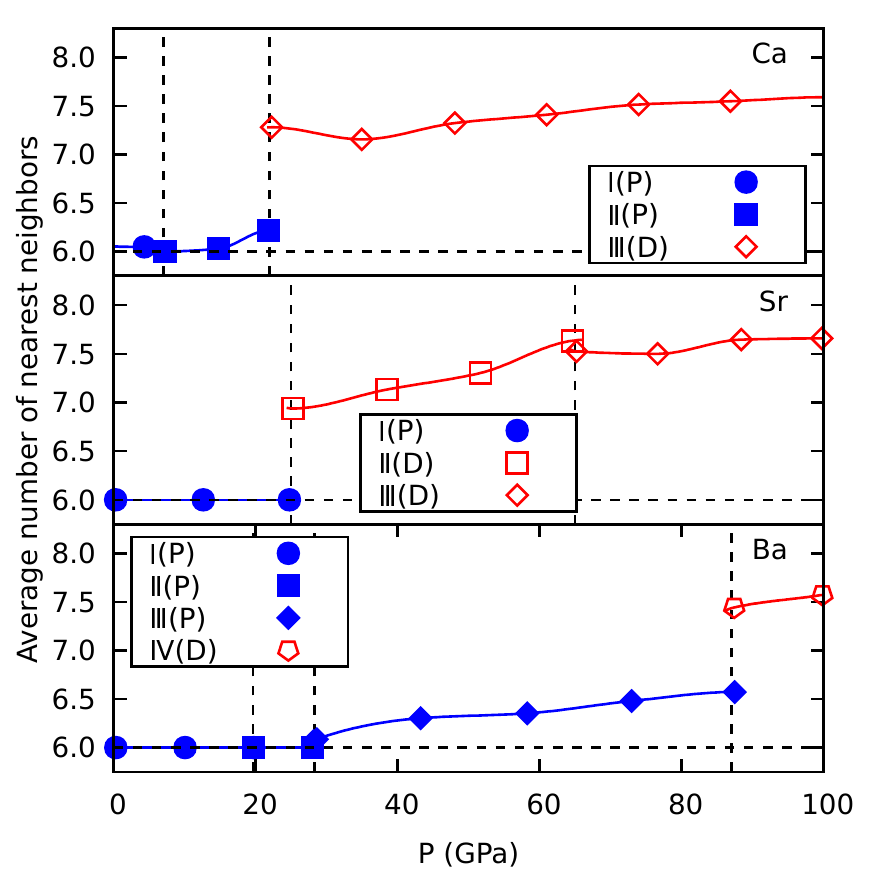}
 \caption{(\textit{color online}) Average number of Bi nearest neighbors as a function of pressure.
 The meaning of the symbols is the same as in Fig.~\ref{fig:Gaps}.}
 \label{fig:Neighbors}
\end{figure}

The neighbors analysis was performed with the help of the
\texttt{Chemenv}\cite{chemenv} module from the \texttt{Pymatgen}\cite{pymatgen,pymatnote} package.
\texttt{Chemenv} provides routines to obtain the best fit of the coordination environment polyhedron for a specific atom and this information was used to estimate  the number of oxygen neighbors for each bismuth atom in Fig.~\ref{fig:Neighbors}.

\begin{acknowledgments}
We acknowledge funding from the Austrian Science Fund FWF through SFB ViCoM, Project F41-P15 and 
computational resources from the VSC3 of the Vienna University of Technology and from the HPC TU Graz.
\end{acknowledgments}

\clearpage

\clearpage

\clearpage


\begin{thebibliography}{52}
\bibitem [{clu()}]{clustered}%
  \BibitemOpen
  \href@noop {} {}\bibinfo {note} {HSE on top of PBE structure describes
  clustered \bbo\ as a semimetal. However, if a proper geometry optimization on
  HSE level is performed, the structure is insulator. Due to computational cost
  and consistency reason we have not performed this optimization for the whole
  pressure range, but only checked the behaviour at a specific
  pressure.}\BibitemShut {Stop}%
\bibitem [{bad()}]{bader_comment}%
  \BibitemOpen
  \href@noop {} {}\bibinfo {note} {In Bader analysis atomic charges are
  separated by ``zero-flux'' surfaces: these are surfaces where there is no
  flux in the gradient vector field of charge density.}\BibitemShut {Stop}%
\bibitem [{pym()}]{pymatnote}%
  \BibitemOpen
  \href@noop {} {}\bibinfo {note} {\texttt{Pymatgen} is a \texttt{Python}
  programming language library for materials analysis that has capabilities of
  generation of phase diagrams, Pourbaix diagrams, diffusion analysis,
  reactions, electronic structure analysis etc. Also it provides tools to
  construct, manipulate and analyze structures. One of them is \texttt{Chemenv}
  module that is developed for chemical environment analysis.}\BibitemShut
  {Stop}%
\end{thebibliography}
\end{document}